\pdfoutput=1
\documentclass[floatfix,showkeys,reprint,amsmath,amssymb,aps,prb]{revtex4-1} 
 \usepackage{float}
\usepackage{bbold}
\usepackage{amsmath}
\usepackage{amssymb}
\usepackage{graphicx}
\usepackage[caption=false]{subfig}
\usepackage{hyperref}
\usepackage{xcolor}
\usepackage[normalem]{ulem}

\makeatletter
\pdfpageheight\paperheight
\pdfpagewidth\paperwidth
\makeatother

\begin{document}
\title{Super-resolving the Ising model with convolutional neural networks}
\author{Stavros~Efthymiou}
\author{Matthew J.~S.~Beach}
\author{Roger G.~Melko}
\affiliation{Perimeter Institute for Theoretical Physics, Waterloo, Ontario N2L 2Y5, Canada}
\affiliation{Department of Physics and Astronomy, University of Waterloo, Waterloo N2L 3G1, Canada}

\date{\today}
\begin{abstract}
    Machine learning is becoming widely used in condensed matter physics. Inspired by the concept of image super-resolution, we propose a method to increase the size of lattice spin configurations using deep convolutional neural networks.
    Through supervised learning on Monte Carlo (MC) generated spin configurations, we train networks that invert real-space renormalization decimations.
    We demonstrate that super-resolution can reproduce thermodynamic observables that agree with MC calculations for the one and two-dimensional Ising model at various temperatures.
    We find that it is possible to predict thermodynamic quantities for lattice sizes larger than those used in training by extrapolating the parameters of the network.
    We use this method to compute the critical exponents of the 2D Ising model, finding good agreement with theory.
\end{abstract}
\maketitle

\section{Introduction}\label{sec:intro}

A primary challenge in the field of quantum many-body physics is the efficient computational simulation of systems with a large number of particles.
Such simulations are crucial for the investigation of strongly-interacting systems, the discovery of exotic phases of matter, and the design of new quantum materials and devices.
Recently, it has been proposed to treat the many-body problem as data-driven, whereby the large dimensionality of the data motivates
the adoption of machine learning algorithms.\cite{CarrMelkoPhases, CarleoRBM, 2018arXiv180800479Z}

In condensed matter systems, neural networks were first used as supervised classifiers that distinguish phases and identify phase transitions, even in unconventional cases when there is no underlying order parameter.\cite{CarrMelkoPhases, WangPT, FermionPhases, BeachXY,extra}
Furthermore, unsupervised generative models have been shown to successfully capture thermal distributions.\cite{TorlaiRBMThermal, IsingGAN, VAE_Ising}
In the quantum case, neural networks are being employed as representations for many-body wavefunctions, with broad applications such as variational ansatz~\cite{CarleoRBM}, guiding functions~\cite{inack_projective_2018},
or for quantum state tomography.~\cite{RBMTomography, Latent, QuCumber,reconstruct}

Early connections between statistical physics and machine learning drew parallels to the renormalization group~\cite{CardyRenorm, KadanoffRenorm}
(RG), a canonical paradigm in physics that involves iteration through a
series of coarse-graining and rescaling procedures.
The mathematical similarity of the RG procedure to the processing of information in multi-layer neural networks has driven
interest in examining the theoretical underpinnings of deep learning.\cite{BenyRG_DLR, MehtaRG}
Conversely, relations between the RG and machine learning have proven useful for physics itself where neural networks have been proposed as generative models that assist RG procedures,\cite{NNRG_Wang} or for identifying relevant degrees of freedom to
decimate.\cite{MI_Maciej, OptRenorm} 
Furthermore, direct applications of neural networks on physical configurations can produce RG or inverse RG flows.\cite{OptRenorm, IsoRG}

Outside of physics, an area of expanding application for machine learning is
image super-resolution, where the goal is to increase the number of pixels in an image while
(subjectively) maintaining the perceivable quality.\cite{srcnn,srgan, FSRCNN, zhao}
Remarkable progress has been made with convolutional neural networks (CNNs), which can be used
to reconstruct high-resolution images to photo-realistic quality.\cite{srgan,zhao}

In this paper, we investigate whether super-resolution methods may be useful in condensed matter and statistical physics by allowing one to produce lattice configurations of larger sizes directly from those obtained for smaller systems.
For concreteness, we focus on the classical Ising model in one and two dimensions.
Our method takes a configuration of Ising spins on a lattice, and subjects it to a
majority rule block spin RG procedure.\cite{InvRG_PRL}
A CNN is then trained to invert this transformation by being exposed to both the higher and lower resolution lattices.
Since some information is necessarily lost in
the RG step, the network output is interpreted as a probabilistic image that is sampled to produce super-resolved images.

We give numerical evidence that the trained super-resolution network performs a probabilistic inverse of the RG transformation and
reproduces thermodynamic quantities on a larger lattices starting only from smaller ones.
In addition, we propose a way to extrapolate the weights of a trained CNN
to apply it to sizes larger than available in the training data.
Using this idea iteratively, we acquire configurations of increasing size that we then use to estimate the critical exponents of the 2D Ising universality class,
obtaining agreement with known theoretical results.

\section{Super-resolution and RG}\label{sec:sr_and_training}

Super-resolution is defined as a mapping $\mathcal{SR}:\mathbb{Z}_2^{L\times L}\rightarrow \mathbb{Z}_2^{fL\times fL}$ from a low-dimensional space of $L\times L$ images to a high-dimensional space of $fL\times fL$ images where $f>1$ is the upscaling factor.
We use $\mathbb{Z}_2$ to denote a binary variable, with $\mathbb{Z}_2^{L\times L}$ denoting an $L \times L$ matrix of binary values.
The objective of image super-resolution in computer vision is to achieve a high perceived quality on the super-resolved image.
This is generally a subjective criterion.
To give a more quantitative definition, quantities like peak signal-to-noise ratio (PSNR) or structural similarity (SIM) have been used.\cite{srcnn,srgan}
However, even such quantities might not be a reliable estimator of the actual quality as perceived by a human.\cite{srgan}

In contrast, statistical physics provides a well-defined objective for super-resolution of physical systems since super-resolved configurations should follow a specific statistical ensemble. Basic thermodynamic quantities like the magnetization or energy serve as an indicator of whether a super-resolved image is consistent with this ensemble.

We proceed by reviewing the real-space decimation of the Ising model before discussing the relevant network architecture.

\subsection{Decimation of the Ising model}
Let $\boldsymbol\sigma \in \mathbb{Z}_2^{2L \times 2L}$ be an Ising configuration of $N$ spins that follows the Boltzmann distribution
\begin{equation}\label{eq:boltzmann_distribution}
    P_K(\boldsymbol\sigma )=\frac{1}{Z}e^{-H(\boldsymbol\sigma)/T}=\frac{1}{Z}e^{K\sum _{\left \langle ij\right \rangle}\sigma _i\sigma _j},
\end{equation}
where $K \equiv 1/T$ and $Z(K)=\sum _{\{\boldsymbol\sigma \}}e^{-H(\boldsymbol\sigma )/T}$ is the partition function. We take periodic boundary conditions (PBC) and include only nearest-neighbor interactions throughout this paper.

Consider the real-space course-graining of the 2D Ising model according to the majority rule.\cite{CardyRenorm} A $2L \times 2L$ lattice is divided into $2\times 2$ blocks where each block is transformed to a spin with the same state as the majority of spins in the block.
If the total sign is zero, we take the sign of the upper left spin to make the procedure deterministic,
instead of the more common probabilistic approach of taking a random sign.\cite{Swedsen_MCRG_PRL}

A low-resolution configuration, $\mathbf{s} \equiv \mathcal{MR}(\boldsymbol\sigma )\in \mathbb{Z}_2^{L\times L}$, is obtained upon applying the deterministic majority rule. Such a configuration follows the marginalized distribution:
\begin{equation}\label{eq:marginalization}
    \tilde{P}_{\tilde K}(\mathbf{s}) = \sum _{\{\boldsymbol\sigma \}}k(\mathbf{s},\boldsymbol\sigma )P_K(\boldsymbol\sigma ),
\end{equation}
where $k(\mathbf{s},\boldsymbol\sigma )$ is the kernel of the transformation.\cite{KardarStatFields}
In this way, the distribution of low-resolution configurations, $\tilde P_{\tilde K}(\mathbf s)$, is directly related to the distribution of high-resolutions $P_K(\boldsymbol\sigma)$.
Evidently, any super-resolution procedure must satisfy the identity relation
\begin{equation}\label{eq:sr1}
    \mathcal{MR}(\mathcal{SR}(\mathbf{s})) = \mathbf s \, .
\end{equation}
A stronger requirement, and the main challenge, is to discover a map such that $\mathcal{SR}(\mathcal{MR}(\boldsymbol\sigma))$ obeys the correct Boltzmann distribution. We emphasize that $\mathcal{SR}(\mathcal{MR}(\boldsymbol\sigma))$ is not necessarily equal to $\boldsymbol \sigma$ since only the distributions need to match, not each individual configuration. Furthermore, in our majority rule decimation, we have not rescaled the Hamiltonian couplings as required in the conventional definition of a complete RG step. It may be possible to learn the rescaling with a neural network; however, in this paper we simply numerically fit the couplings as needed for a consistent rescaling.

\subsection{Network architecture}

We now attempt to invert the majority rule procedure using a supervised learning approach.
The unknown super-resolution mapping $\mathcal{SR}$ is parametrized with a deep convolutional neural network (CNN).
CNNs are ideal for our problem due to their utility in image processing tasks. Moreover, the weight-sharing property of convolutions allows extrapolation to larger sizes.

\begin{figure}[t]
    \includegraphics{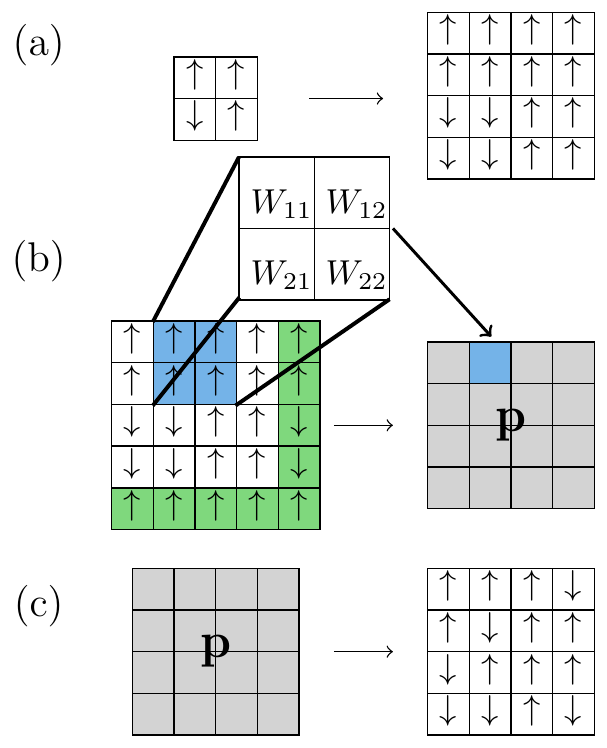}
    \caption{(a) Upsampling by replacing each up (down) spin with a block of four up (down) spins. (b) The weights $W$ convolve local regions together and add a bias $b_i$. Applying a sigmoid function elementwise gives the probabilities of each site being up as $\mathbf{p}$. Green sites correspond to the PBC padding. (c) Sampling $\mathbf{p}$ gives discrete Ising spins on the super-resolved $2L \times 2L$ lattice.} \label{fig:network_description}
\end{figure}

The first layer of our network is an upsampling layer that increases the resolution from $L\times L$ to $2L \times 2L$ by transforming each up (down) spin to a block of four up (down) spins (Fig.~\ref{fig:network_description}a).
At very low temperatures configurations are fully polarized and we expect this upsampling to be highly accurate.
However, non-fully-polarized blocks appear at higher temperatures, making naive upsampling insufficient.
In order to alleviate this, the convolution layers that follow must add the required statistical fluctuations, similar to the Monte Carlo sweep procedure in Ref.~[\onlinecite{InvRG_PRL}].

Each convolution layer takes a configuration $\mathbf{x}\in \mathbb{R}^{2L \times 2L}$ as input and applies the transformation $f(W*\mathbf{x} + \mathbf{b})$, where $*$ denotes the convolution operation, ${W}$ is the so-called filter, $\mathbf{b}$ is a bias vector and $f$ is a non-linear differentiable function applied element-wise (Fig.~\ref{fig:network_description}b).
This function is known as an activation function and the typical choice is the rectified linear function $\mathrm{ReLU}(x)=\max (0, x)$.
The effect of each convolution is to combine local features within a $n_f \times n_f$ region (filter size). 
A consequence of this is that each convolution layer reduces the images size by eliminating the right-most and bottom edges.
To avoid truncating the image edge, we surround the original configurations with additional spins from the periodic boundary conditions (Fig.~\ref{fig:network_description}b).
This has the advantage of respecting the boundary conditions of the underlying physical model.

In order to obtain a discrete Ising configuration, the sigmoid activation $\sigma (x) = 1 / (1 + e^{-x})$ is used in the final layer, giving an output $\mathbf{p}\in [0, 1]^{2L \times 2L}$.
This is treated as the probability that the corresponding spin is up.
A discrete configuration is then obtained by sampling $\mathbf{p}$ for each lattice site (Fig.~\ref{fig:network_description}c).

The network is characterized by parameters $\theta$, which include all the weights, $W$, and biases, $\mathbf{b}$, from each convolution layer.
These are tuned to minimize a loss function defined on a dataset of inputs $\mathbf s_i\in\mathbb{Z}_2^{L\times L}$ and targets $\boldsymbol{\sigma}_i\in\mathbb{Z}_2^{2L \times 2L}$ with $i \in \{1, 2, \ldots , n\}$, where $n$ is the number of samples in the dataset.
In contrast to typical supervised learning applications (e.g.~handwritten digit recognition), here the dimensionality of the output is larger than the input.
The loss quantifies the distance between predicted output $\mathcal{SR}_{\theta}(\mathbf s_i)$ and the original high-resolution $\boldsymbol \sigma_i$.
Minimization is done with back-propagation~\cite{backpropagation} which involves calculations of gradients, and thus cannot be done using the final sampled (discrete) output.
To be consistent with the interpretation of the continuous outputs $\mathbf{p}$ as probability we use the cross-entropy loss function;
\begin{align}\label{eq:cross_entropy_loss}
    \mathbb{L}(\{\boldsymbol \sigma_i\}, \{\mathbf p_i \})
     & = -
    \sum _{i=1}^n
    \left[ \boldsymbol \sigma_i \cdot \ln \mathbf{p}_i + \left (1-\boldsymbol \sigma_i \right) \cdot \ln \left (1-\mathbf{p}_i\right )\right ],
\end{align}
where $i\in \{1,2,\ldots,n\}$ and $\cdot$ denotes the element-wise product between matrices.
Note, we are free to add additional terms to Eq.~(\ref{eq:cross_entropy_loss}) to assist training. 
For example, as we will see, it is sometimes beneficial to introduce a term
proportional to $|E(\boldsymbol\sigma_i) - E(\mathbf p_i)|^2$.

\subsection{Extrapolation to larger lattices }\label{sec:weight_extrapolation}
We approximate the super-resolution mapping $\mathcal{SR}$ as a neural network $\mathcal{SR}_{\theta}$ with parameters $\theta$.
In order to train the network $\mathcal{SR}_{\theta}$, we need access to $2L \times 2L$ configurations. We obtain these with Monte Carlo simulations.
In this sense,
the method does not allow us to access sizes larger than the ones we have already simulated.
It would be useful if super-resolution could be used to access sizes that cannot be obtained by other means.
We propose a simple method to do precisely this by exploiting the weight sharing property of convolutions.
Namely, the size of the weight matrix $W$ (and the bias vector $\mathbf{b}$) on a convolutional layer is independent of the input and output size.

In order to apply the convolution to a larger image, we only have to ``slide'' (Fig.~\ref{fig:network_description}b) the trained matrix $W$ over a larger surface.
The first upsampling layer does not contain any weights and can be trivially applied to any size.
Therefore, using the weights of the trained $\mathbb{Z}_2^{L\times L}\rightarrow \mathbb{Z}_2^{2L \times 2L}$ network, we can define a new $\mathbb{Z}_2^{L'\times L'}\rightarrow \mathbb{Z}_2^{2L'\times 2L'}$ network that can be used to double any input size $L'$.

\begin{figure}[tb]
    \centering
    \includegraphics[width=0.48\textwidth]{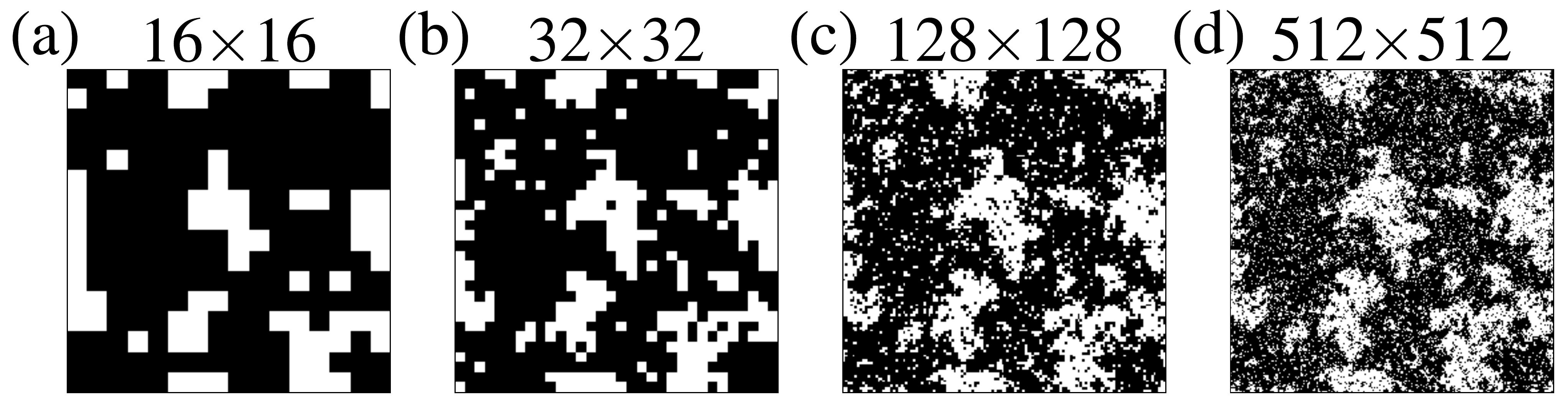}
    \caption{\label{fig:confs_crit}Critical configurations obtained using the weight extrapolation idea presented in Section~\ref{sec:weight_extrapolation}. We show the original Monte Carlo configuration in (a) and the results after (b) one, (c) three and (d) five consecutive super-resolutions.}
\end{figure}

Accordingly, we can take the $2L \times 2L$ output of the first super-resolution as the input of a new network $\mathbb{Z}_2^{2L \times 2L}\rightarrow \mathbb{Z}_2^{4L\times 4L}$.
Doing this repeatedly, we can create a chain of increasing sizes $(\mathbb{Z}_2^{16 \times 16}\rightarrow \mathbb{Z}_2^{32 \times 32} \rightarrow \mathbb{Z}_2^{64 \times 64} \rightarrow \dots)$.
In this chain, only the smallest configurations are generated with MC, while the rest are result of successive super-resolutions with the same $\theta$ parameters.
Figure~\ref{fig:confs_crit}a shows a configuration of the 2D Ising model at criticality, while Fig.~\ref{fig:confs_crit}b,c,d show the super-resolutions obtained from this configuration.

In summary, we use configurations 
$\boldsymbol{\sigma}_i\in \mathbb{Z}_2^{2L\times 2L}$ 
to generate decimated configurations 
$\mathcal{MR}(\boldsymbol{\sigma}_i)\in \mathbb Z_2 ^{L\times L}$.
We then train a CNN network $\mathcal{SR}_{\theta}$ to invert this transformation.
If the network is trained correctly, new configurations 
$\mathcal{SR}_{\theta}(\mathcal{MR}(\boldsymbol{\sigma}))$, 
should obey the Boltzmann distribution within reasonable error.

In the following sections, we test this procedure by calculating observables in the 1D and 2D Ising models.

\section{One-dimensional Ising model}\label{sec:results1d}
In this section, we show the results of the super-resolution scheme applied to the 1D Ising model. For this model, the Hamiltonian is self-similar under RG steps and the decimation is exactly solvable. This serves as a useful benchmark to test the validity of our super-resolution scheme before assessing the more interesting 2D case.

\begin{figure*}[bhtp]
    \centering
    \includegraphics[width=1.0\textwidth]{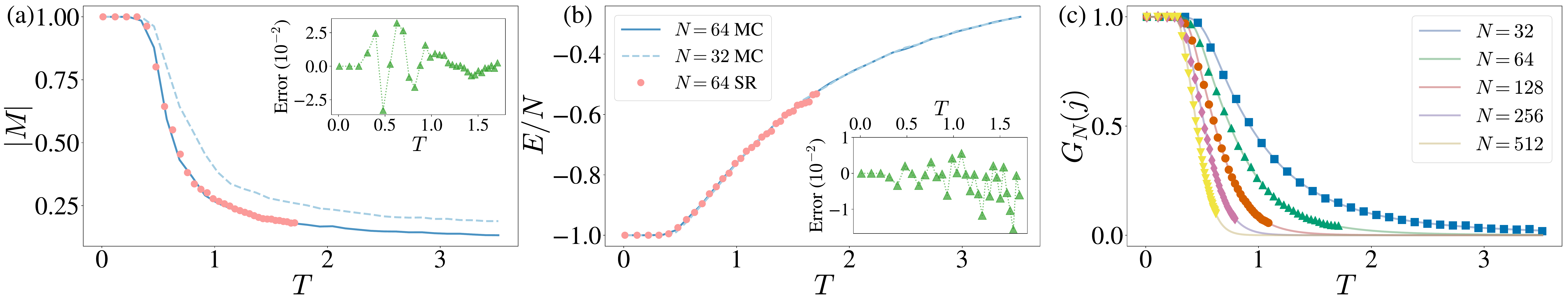}
    \caption{\label{fig:ups_real1D}(a)
    Absolute magnetization and (b) energy per spin for the 1D Ising model.
    We denote Monte Carlo results at low ($N=32$) and high ($N=64$) resolution with MC\@.
    The super-resolution (SR) results were obtained by using the $N=32$ MC data as input to an $\mathbb{Z}_2^{16}\rightarrow \mathbb{Z}_2^{32}$ network, and extrapolating new $N=64$ configurations.
    SR temperatures are adjusted according to the inverse of Eq.~(\ref{eq:renormalized_temp_1d}).
    This shrinks the temperature range as the inverse RG transformation flows towards $T=0$.
    Inset plots correspond to the error between SR predictions and MC results.
    (c) Two-point function of the 1D Ising model with $j=N^{0.8}/5$. Solid lines correspond to Eq.~(\ref{eq:tpf_theory_1d}) and marked points to the super-resolution prediction.
    We use MC data for $N=32$, while all other sizes are obtained from consecutive super-resolutions.
    }
\end{figure*}

We begin with training the network with two lattice sizes, $N$ and $2N$, before attempting extrapolation to larger sizes.
We use a dataset consisting of temperatures ranging from $T=0.01$ to $T=3.5$ with $N=32$ spins.
At each temperature, we create training and testing sets consisting of $n=10^4$ configurations generated via standard Monte Carlo.
Instead of the majority rule, we use real-space block-spin decimation to obtain a $N=16$ chain.\cite{KadanoffRenorm}
For each temperature, we implement a different network and optimize using Adam,\cite{AdamOptimizer} with a batch size of $10^3$.
Instead of training for a specific amount of epochs, we cease training when the validation loss stops improving using early stopping.

We find that the network achieves reasonable accuracy at each temperature when evaluated in the test set (Fig.~\ref{fig:test_1D} in Appendix~\ref{sec:app_test}).
This shows that the network performs an approximate inverse of the block-spin decimation.
We note that this inversion is demonstrated only at the level of thermodynamic observables and not the whole statistical ensemble.
In the remainder of this section, we will focus on generating larger sizes than present in the training set.

So far, we have not discussed the temperature of configurations or the rescaling part of an RG step, but it will be essential to obtaining larger configurations.
Conveniently, in 1D we can exactly calculate the marginalization of Eq.~(\ref{eq:marginalization}).
The Hamiltonian is self-similar under the RG decimation with the rescaled couplings
\begin{equation}\label{eq:renormalized_temp_1d}
    \tilde{K} = f(K) = \frac{1}{2} \ln \cosh 2K \, .
\end{equation}
Thus, under successive RG steps, the temperature $T$ of a configuration flows towards infinity.

In effect, we train the network to take configurations 
$\mathbf{s}_i$ at couplings $\tilde K$
(obtained by applying
$\mathcal{MR}$
to $\boldsymbol{\sigma}_i$ at $K$)
and then super-resolve to a high-resolution configuration
$\mathcal{SR}(\mathbf{s}_i)$ at $K$.
If we use this configuration as the new input to the network, we can generate new, larger configurations indefinitely,
following the method described in Section~\ref{sec:weight_extrapolation}.
After one super-resolving step, the new configurations produced by the network are
$\mathcal{SR}(\mathcal{SR}(\mathbf{s}_i))$ at  $f^{-1}(K) = f^{-1}(f^{-1}(\tilde K))$.
Here we see that is important to know how to apply the rescaling step (\ref{eq:renormalized_temp_1d}).

To validate our extrapolation proposal, we compare the magnetization and energy of super-resolved configurations with Monte Carlo results. Fig.~\ref{fig:ups_real1D} shows the results of the network after adjusting the temperature with Eq.~(\ref{eq:renormalized_temp_1d}).
We stress that the network was trained on $\mathbb{Z}_2^{16} \rightarrow \mathbb{Z}_2^{32}$ data, yet predicts $N=64$ accurately by extrapolation.

As another test of the super-resolving network, we repeat extrapolation up to $N=512$ spins and calculate the two-point function,
$G_N(j)=\left \langle \sigma _1\sigma _{1+j} \right \rangle$, for configurations from each generated size.
In 1D, the exact value for this quantity is:\cite{Baxter}
\begin{equation} \label{eq:tpf_theory_1d}
    G_N(j;K)=\frac{\tanh ^jK + \tanh ^{N-j}K}{1+\tanh ^NK} \, .
\end{equation}
In Fig.~\ref{fig:ups_real1D}c we plot the two-point function for the different sizes with $j=N^{0.8}/5$.
We note that the choice $j=N^{0.8}/5$ does not have a particular physical significance as it is possible to obtain similar accuracy for different choices of $j$.

Generally, errors are expected to increase with each super-resolution step. However, in the current case, the temperature flows towards zero under the inverse Eq.~(\ref{eq:renormalized_temp_1d}), so the extrapolation scheme remains stable even after multiple super-resolutions.

We have demonstrated numerically with two different methods that our super-resolution mapping can successfully capture thermodynamic quantities, as an approximate inverse RG transformation.
The network parameter extrapolation is particularly effective in 1D where we know exactly how to rescale the temperature from Eq.~(\ref{eq:renormalized_temp_1d}).

\section{Two-dimensional Ising model}\label{sec:results2d}

Following the success of the 1D case, we proceed by training a 2D, $\mathbb{Z}_2^{8 \times 8}\rightarrow \mathbb{Z}_2^{16 \times 16}$ network using Monte Carlo generated datasets. The deterministic majority rule is now used instead of simple decimation.
We again find success on the testing set (Appendix~\ref{sec:app_test}) and seek to explore extrapolation to larger sizes.

\begin{figure*}[ht]
    \centering
    \includegraphics[width=0.95\textwidth]{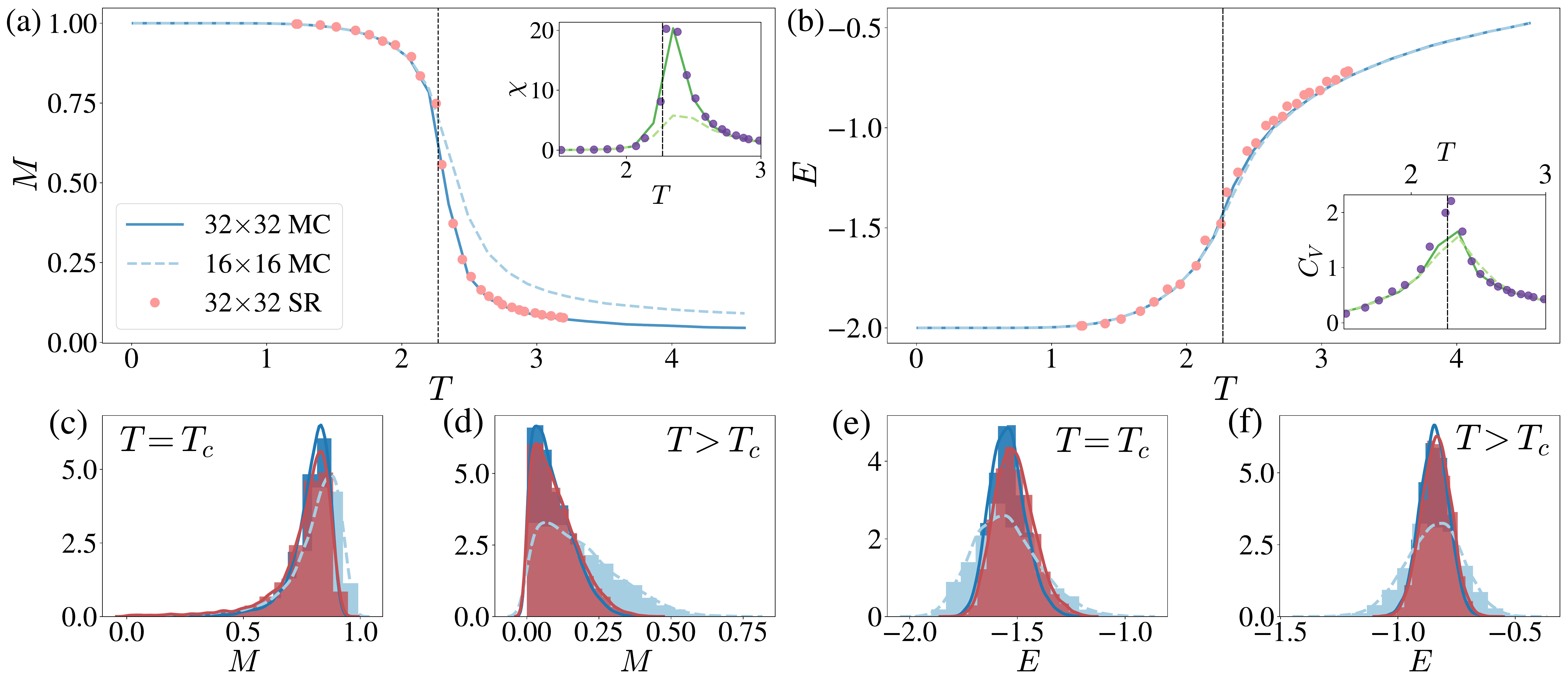}
    \caption{\label{fig:ups_real2D}(a) Absolute magnetization with the susceptibility and (b) energy with specific heat for the 2D Ising model. MC denotes Monte Carlo results while SR is obtained by super-resolving the $16\times 16$ MC configurations using the extrapolation of the $\mathbb{Z}_2^{8 \times 8}\rightarrow \mathbb{Z}_2^{16 \times 16}$ network. SR temperatures were rescaled using the numerical transformation, which shrinks temperature range towards criticality.
    (c), (d) Probability distributions of magnetization at $T=2.2010\simeq T_c$ for and $T=2.9313>T_c$, and (e), (f) the probability distributions for energy.}
\end{figure*}

The main challenge in 2D is that the marginalization of Eq.~(\ref{eq:marginalization}) cannot be done analytically and hence we cannot simply rescale using Eq.~(\ref{eq:renormalized_temp_1d}). The 2D model is not self-similar under the block-spin RG transformation, as the Hamiltonian that corresponds to the decimated distribution contains interactions beyond nearest-neighbors.\cite{Swedsen_MCRG_PRB}

To obtain a 2D analogy of Fig.~\ref{fig:ups_real1D}, we approximate the temperature correction numerically for observables.
In order to numerically find the transformation $f^{-1}:\tilde{K}\rightarrow K$ we compare observables calculated on an $8\times 8$ MC configuration with those calculated on an $8\times 8$ decimated one.
We require that the corresponding curves collapse when the transformation is applied to the MC data. This procedure is discussed in detail in Appendix~\ref{sec:app_rescale}.
We note that this rescaling procedure has no direct physical interpretation since the 2D Ising Hamiltonian is not self-similar after a block spin RG transformation, and therefore temperature alone is not sufficient to describe the coupling space of the RG configuration.
Thus, here we use this procedure only to demonstrate that our results are consistent with the inverse-RG nature of super-resolution.

\subsection{Thermodynamic observables}\label{sec:thermo2d}

We proceed by extrapolating the parameters of the trained $\mathbb{Z}_2^{8 \times 8}\rightarrow \mathbb{Z}_2^{16 \times 16}$ network and using it to super-resolve $16\times 16$ MC configurations to $32\times 32$.
We present the results for magnetization and energy in Fig~\ref{fig:ups_real2D}a,b.
As in 1D, we see that the rescaling makes predicted SR observables match the MC results, indicating again that the network performs an approximate inversion of the RG transformation as desired.

We note that at high temperatures, the noise is largely random and difficult to learn.
In contrast to 1D, we add a regularization term in the loss function, which compares the energy of the super-resolved configuration to that of the original one. This does not use any more information than already present in the training data, but results in better convergence of the network for high temperatures.

To corroborate our findings, we calculate the probability distributions of magnetization and energy in Fig.~\ref{fig:ups_real2D}, for $T\simeq T_c$ and $T >T_c$.
We see that super-resolution captures not only the average values of magnetization and energy, but their entire probability distribution.

We expect that by increasing the extrapolation to larger sizes, any error in the original data or imperfections in the network will propagate.
However, we suggest that for a single extrapolation, the network does remarkably well.
Even multiple extrapolations may serve as a good starting point for large-scale simulations, e.g.~possibly shortening equilibration time in a Monte Carlo procedure.

\begin{figure}[htbp]
    \centering
    \includegraphics[width=0.9\columnwidth] {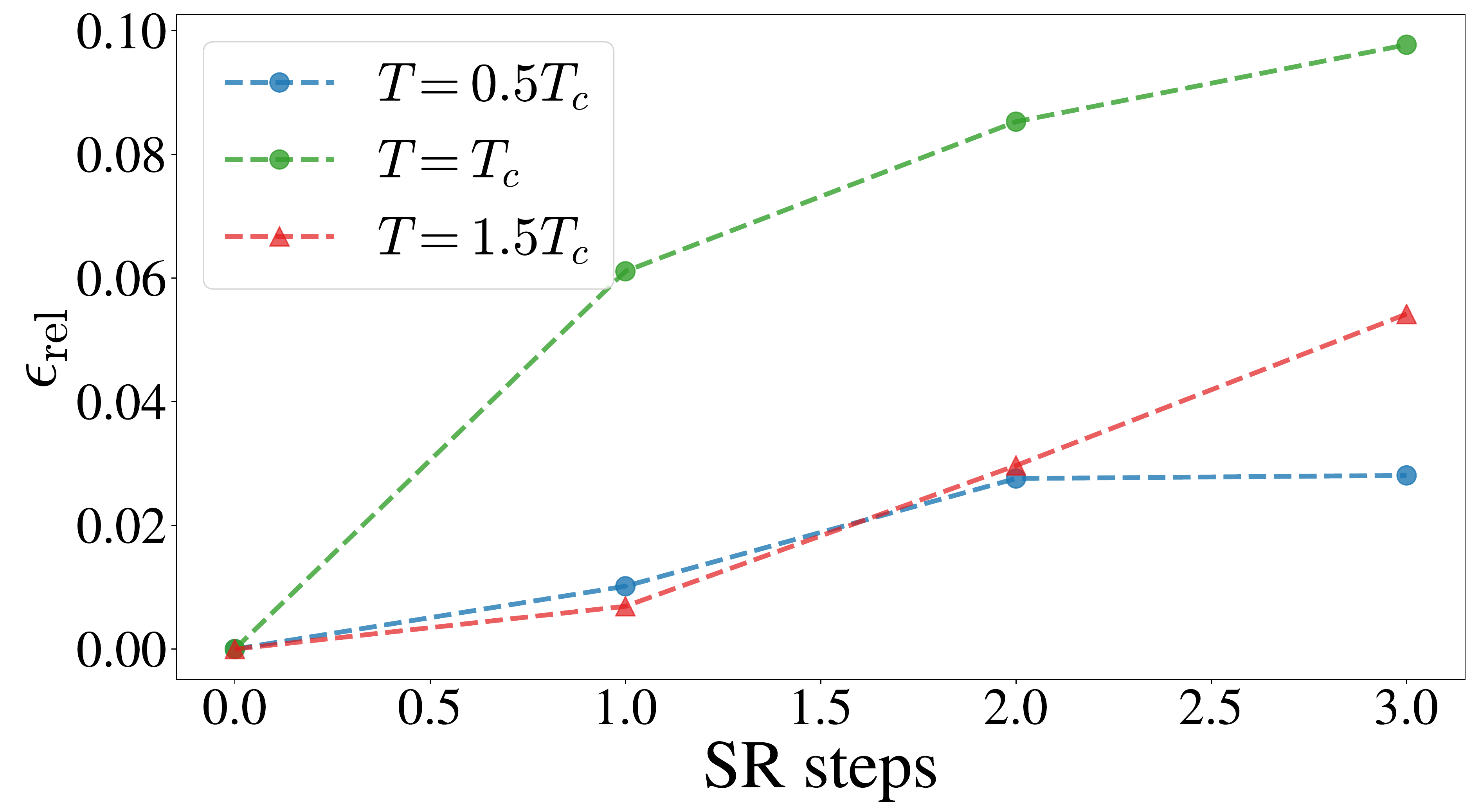}
    \caption{\label{fig:error} The relative error in the energy $\epsilon_{\rm rel} = |E_{\rm MC} - E_{\rm SR}| / E_{\rm MC}$ as a function of number of SR steps. At $T=T_c$, the algorithm is the most inaccurate, with around $6\%$ error after one step. }
\end{figure}

A natural question is how the accuracy in thermodynamic observables scales with the extrapolation procedure.
In Fig.~\ref{fig:error} we show the relative error in the energy after repeating the upsampling.
We see here that the error grows with each successive upscaling as expected.
This does not hold for all thermodynamic quantities, as the magnetization typically stays within $0.5\%$ error at $T_c$ for up to three super-resolutions.
Motivated in particular by the accuracy of the magnetization, we consider extracting the critical exponents from finite-size scaling with super-resolved configurations.

\subsection{Critical Exponents}\label{sec:finite_size_scaling}
An interesting application of super-resolution is the calculation of critical exponents.
In principle, we can avoid rescaling if we focus on the fixed point of the RG transformation, where the Hamiltonian is self-similar.
Here we crudely approximate~\cite{InvRG_PRL} the fixed point with the nearest-neighbor Hamiltonian precisely at the critical temperature.
We take $10^5$ samples of $16\times 16$ Monte Carlo configurations and we repeatedly extrapolate to reach sizes up to $128\times 128$.
We stress that Monte Carlo simulation is required only on the smallest size (in our case $16\times 16$) as all larger sizes are obtained by extrapolating.

We use the predicted configurations to calculate the critical exponents for the 2D Ising universality class using the finite-size scaling hypothesis.\cite{SandvikGuide}
According to this, exactly at criticality $\chi \propto L^{\gamma /\nu }$, where $\gamma $ and $\nu $ are the susceptibility and correlation length critical exponents respectively. We can estimate the $\gamma /\nu $ ratio from the slope in a log-log $(L, \chi )$ plot (Appendix~\ref{sec:app_expo}).
Similarly, the two-point function vanishes algebraically $G(r)\sim 1/r^{d-2+\eta }$ at criticality, allowing us to estimate the anomalous dimension $\eta $ from the slope of a log-log $(r, G(r))$ plot. The magnetization exponent $\beta $ can be calculated similarly. The exponents found through this method are presented in Table~\ref{tab:critical_exponents}, where we see the remarkable agreement with analytical results.

\begin{table}[htbp]
    \centering
    \begin{tabular}{ccccc}
        \hline \hline
        Exponent  & Super-resolution   & Error  \\ \hline
        $\beta $  & 0$.1234 \pm 0.006$ & 1.3\%  \\
        $\gamma $ & $1.7544 \pm 0.01$  & 0.25\% \\
        $\eta _1$ & $0.2460 \pm 0.01$  & 1.6\%  \\
        $\eta _2$ & $0.2459 \pm 0.01$  & 1.6\%  \\
        \hline \hline
    \end{tabular}
    \caption{Critical exponents of the 2D Ising universality class. We give the mean and standard error of 60 independent repetitions of training and critical exponent calculation from $16\times 16$ to $128\times 128$. The error is calculated in respect to exact values in the thermodynamic limit.}
    \label{tab:critical_exponents}
\end{table}

\section{Discussion}\label{sec:conclusions}

We have investigated whether super-resolution techniques can be used to successfully increase the size of physical configurations sampled from the 1D and 2D Ising Hamiltonian.
Inspired by recent applications of deep learning, we used a convolutional neural network for this task.
We performed supervised training with a set of Monte Carlo configurations as output, and their corresponding RG-decimated counterparts as input.
Therefore, the network was essentially trained to double the size of configurations by performing a transformation approximately equivalent to an inverse RG step.

Despite the challenge in rigorously defining the inverse RG transformation due to the loss of information during the decimation,
we found that our super-resolution scheme can accurately capture thermodynamic observables over a wide range of temperatures.
We further proposed a method to extrapolate the trained weights and biases, and used them to access arbitrary lattice sizes larger than those used for training.
We found that the extrapolation worked well for both 1D and 2D systems, and we were able to compute critical exponents in the 2D case which show agreement with analytical results to within $2\%$.
To achieve this success, the method hinges on knowing the rescaling of the couplings $\tilde K = f(K)$ at each RG step. We expect that by modifying the network architecture to include an auxiliary parameter, one could learn this rescaling directly from the data, possibly with more accuracy than the approximations here.

Looking forward, these techniques may be beneficial to the large-scale simulation of complex physical systems. For example, our extrapolation may provide approximate initial configurations for further optimization procedures such as Monte Carlo updates.
Decimation and super-resolution could be used to propose non-local updating procedures in models that suffer from long autocorrelation times, such as lattice quantum chromodynamics. \cite{
    shanahan_machine_2018}
Other interesting extensions of the current work could involve systems with disordered couplings, where more sophisticated RG techniques, such as the energy based Ma-Dasgupta-Hu method, may be necessary.\cite{ma_random_1979,
    dasgupta_temperature_1980}

Ultimately, ideas analogous to the weight extrapolation might allow one to generate approximate configurations for lattice sizes that are inaccessible by other means.
A quantum generalization could prove particularly useful as a way to generate approximate configurations that are beyond reach of current quantum Monte Carlo or tensor networks methods.

\begin{acknowledgments}
    The authors would like to thank
    J.~Carrasquilla,
    A.~Golubeva,
    L.~E.~Hayward Sierens,
    B.~Kulchytskyy,
    P.~Ponte,
    I.~Tamblyn,
    and
    G.~Torlai,
    for many useful discussions.
    This research was supported by the Natural Sciences and Engineering Research Council of Canada (NSERC), the Canada Research Chair program, and the Perimeter Institute for Theoretical Physics. This work was made possible by the facilities of the Shared Hierarchical Academic Research Computing Network (SHARCNET) and Compute/Calcul Canada. We also gratefully acknowledge the support of NVIDIA Corporation with the donation of the Titan Xp GPU used for this research. Research at Perimeter Institute is supported by the Government of Canada through Industry Canada and by the Province of Ontario through the Ministry of Research \& Innovation.
\end{acknowledgments}

\bibliography{references}
\newpage
\appendix
\section{Importance of sampling}\label{app:sampling_importance}
As mentioned in Section~\ref{sec:sr_and_training}, the continuous output after the last sigmoid layer is interpreted as the probability that the spin in the corresponding site is up. Therefore, the super-resolved configuration is obtained by sampling this continuous output.
The $\mathcal{F_\theta}:\mathbb{Z}_2^{L \times L}\rightarrow \mathbb{Z}_2^{2L \times 2L}$ mapping consists of the convolutional network and the sampling procedure.

This sampling procedure makes the $\mathcal{SR}$ mapping non-deterministic.
We believe that sampling is crucial for the method to work. Here we give some numerical evidence to corroborate this statement.

The loss of information in the majority rule RG is associated with the different types of blocks that give the same decimated spin.
For example, consider a block with 4-up/0-down spins and one with 3-up/1-down. Both would lead to an up spin in the decimated configuration.
In order to capture the correct thermodynamics, the inverse RG procedure should give these different block types with the correct proportion at each temperature.
To investigate whether this happens, we observe that the type of each block can be uniquely defined by the sum of spins contained in the block. In the $\mathbb{Z}_2=\{0,1\}$ convention, this sum is the number of up spins and goes from 0 to 4.

\begin{figure}[bthp]
    \centering
    \includegraphics[width=0.48\textwidth]{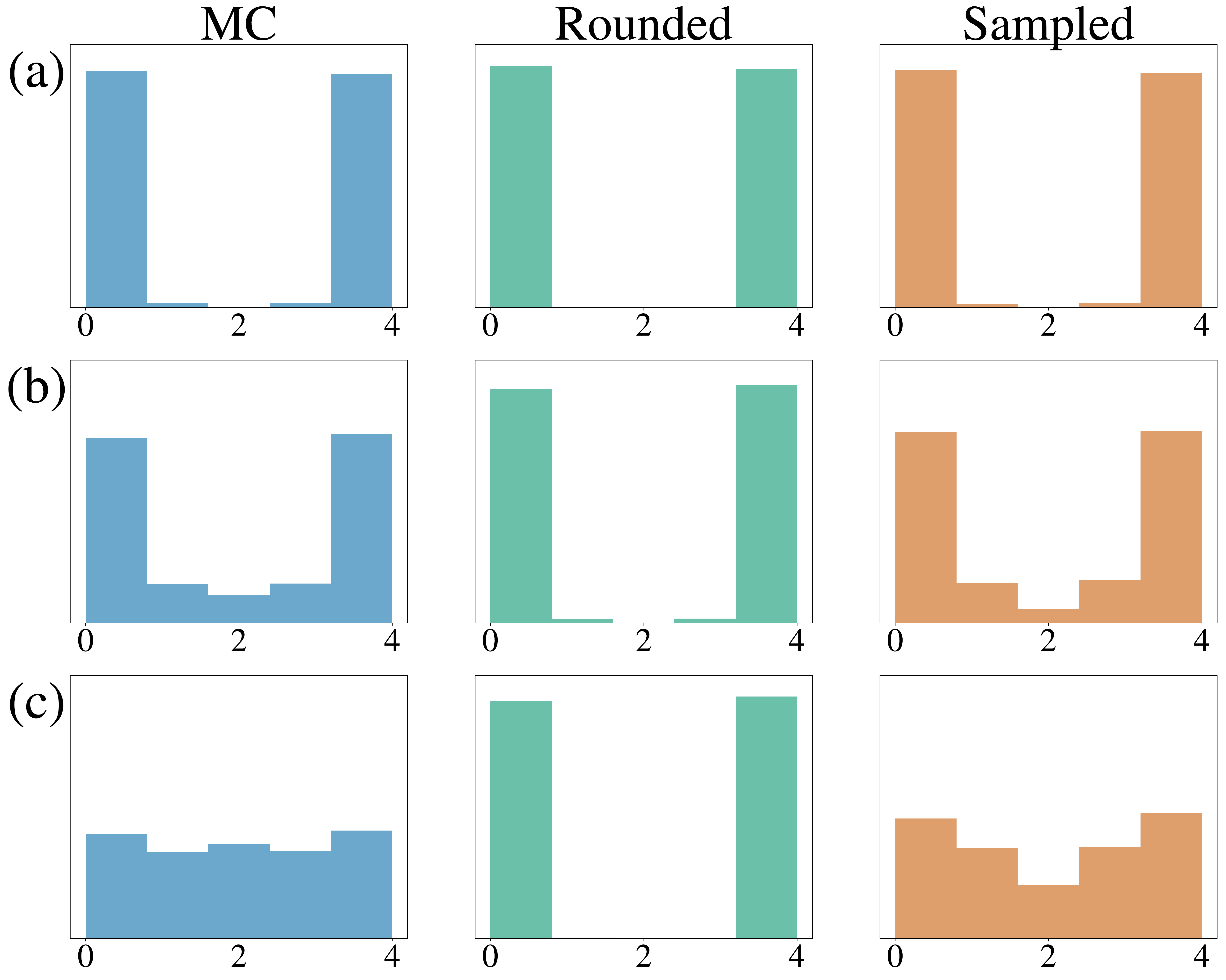}
    \caption{\label{fig:sumhist_samp}Histograms of the different $2\times 2$ block sums at three different temperatures (a) $T=1.4706$ (low), (b) $T=2.2010$ (critical) and (c) $T=2.9313$ (high). The first column corresponds to MC configurations, while the second and third to rounded and uniformly sampled network output respectively.}
\end{figure}

In Fig.~\ref{fig:sumhist_samp} we give the number of appearances of each block sum in the original MC configurations and different interpretations of the network's output (rounding or sampling). The height of each bar is calculated by summing the appearances of each block sum over each configuration. At low temperatures, most configurations are fully polarized with the value 0 (all down) and 4 (all up). In this case, we do not have information loss during RG and there is no significant difference between rounding and sampling the output. At temperatures near and above criticality, non-fully polarized blocks start to appear, increasing the appearance of intermediate sums (1 to 3). Rounding fails to capture the non-fully polarized blocks, making the use of sampling imperative. Even sampling cannot accurately capture the blocks with 2-up/2-down spins, indicating a possible systematic inaccuracy in our method that could be improved in further work.

\section{Testing Data}\label{sec:app_test}

In Sections~\ref{sec:results1d} and~\ref{sec:results2d} we demonstrated that super-resolution can capture the thermodynamics of sizes larger than the ones used in training, simply by extrapolating the network parameters. Before this step, it is important to directly evaluate the network's performance on the objective that it was trained on, namely the super-resolution of down-sampled configurations. We present this evaluation here.

\begin{figure*}[htbp]
    \centering
    \includegraphics[width=1.0\textwidth]{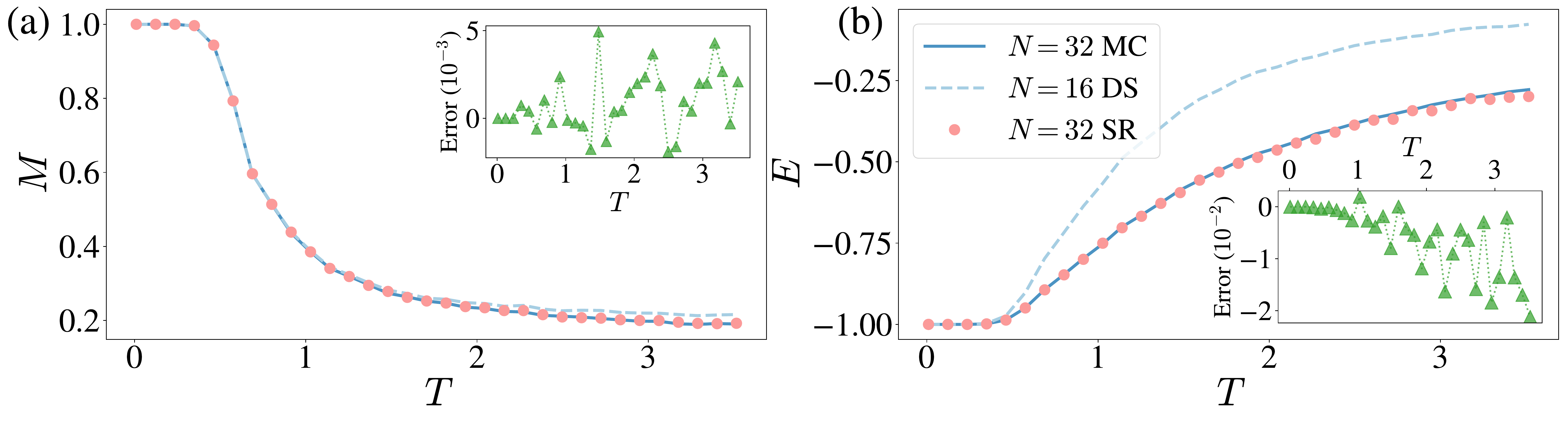}
    \caption{\label{fig:test_1D} (a) Magnetization and (b) energy of the 1D Ising model. The dashed line corresponds to observables computed with the down-sampled (DS) configurations used as the network's input.}
\end{figure*}

\begin{figure*}[htbp]
    \centering
    \includegraphics[width=1.0\textwidth]{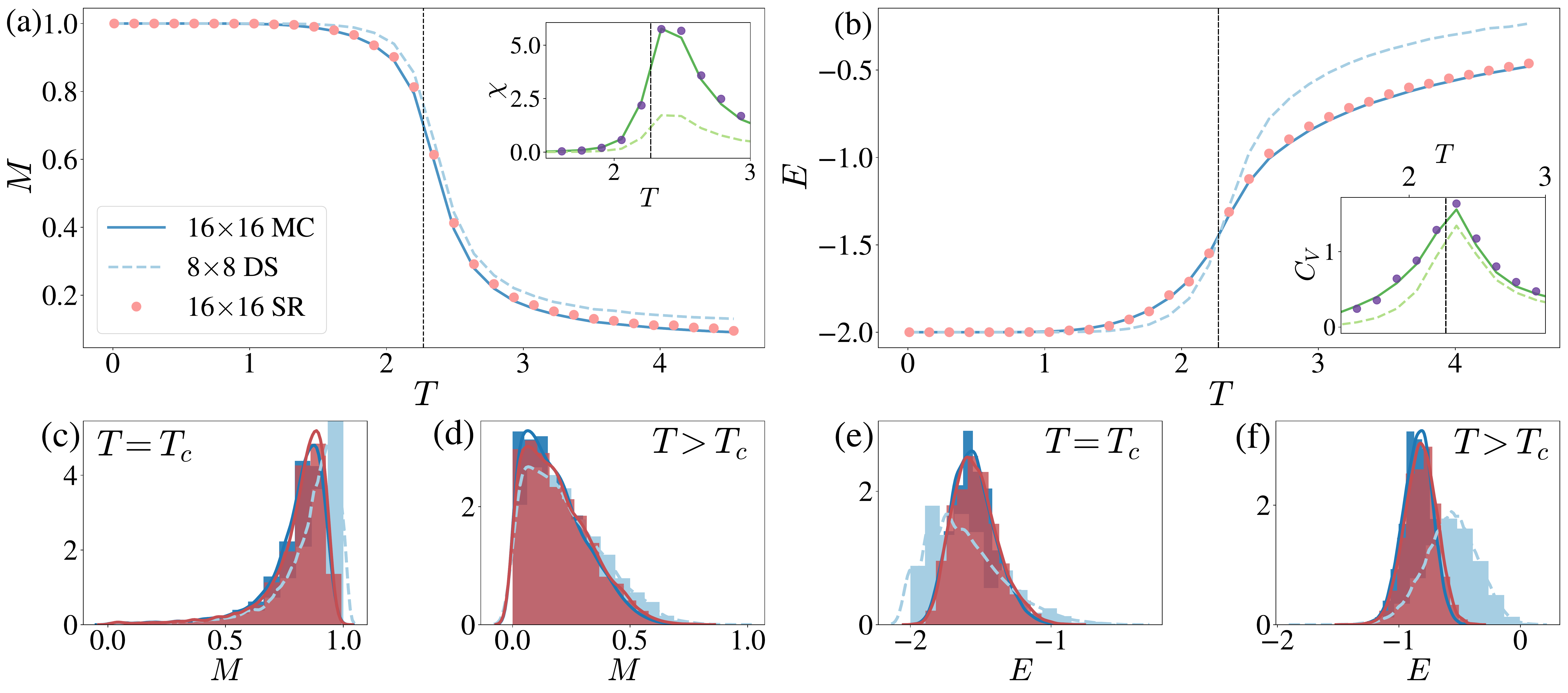}
    \caption{\label{fig:test_2D}(a) Magnetization (with susceptibility) and (b) energy (with specific heat) of the 2D Ising model. MC denotes Monte Carlo results while SR is obtained by super-resolving the $8\times 8$ downsampled (DS) configurations using the $\mathbb{Z}_2^{8 \times 8}\rightarrow \mathbb{Z}_2^{16 \times 16}$ network.
    Below: Probability distributions of magnetization and energy at $T=2.2010\simeq T_c$ for (c, e) and $T=2.9313>T_c$ for (d, f). The observables are binned into 15 bins to obtain these histograms. Colors follow the convention of the plots (a, b).}
\end{figure*}

In Fig.~\ref{fig:test_1D} we test the 1D $\mathbb{Z}_2^{16}\rightarrow \mathbb{Z}_2^{32}$ network on super-resolving down-sampled (DS) $N=16$ configurations and we show that it correctly captures the $N=32$ MC results.
Fig.~\ref{fig:test_2D} gives the same test for the 2D $\mathbb{Z}_2^{8\times 8}\rightarrow \mathbb{Z}_2^{16\times 16}$ network.
As in the main text, 2D results are corroborated by the histograms of Fig.~\ref{fig:test_2D}, which show that the whole magnetization and energy distributions predicted by super-resolution match the corresponding MC distributions.
In each dimension, DS configurations are obtained by applying the respective decimation transformation on the large MC samples.
DS configurations are then used as the network's input for training. Therefore, these results indicate directly that the network approximately inverts the down-sampling procedure, the caveat being the underrepresentation of 2-up/2-down blocks. Note that, unlike the
main text, in the current appendix we did not rescale the temperature of SR predictions. Rescaling is not needed because the input is decimated configurations that follow the marginalized distribution in Eq.~(\ref{eq:marginalization}).

As expected, the errors are typically lower for the direct testing case presented in the current Appendix, compared to the extrapolation results of Section~\ref{sec:results1d}. Also, errors are generally larger at higher temperatures, where thermal noise becomes more random and more difficult to learn.

\section{Approximate rescaling}\label{sec:app_rescale}
As explained in Section~\ref{sec:results2d}, when extrapolating the network parameters, the new larger configurations generated are $\mathcal{SR}(\mathcal{SR}(\mathbf{s}_i)$ at $f^{-1}(K)$. The rescaling function $f^{-1}$ was trivially found in 1D from the known analytical result. However such a result does not exist in the 2D case. Here we describe a method to approximate this rescaling function numerically.

\begin{figure}[thbp]
    \centering
    \includegraphics[width=0.45\textwidth]{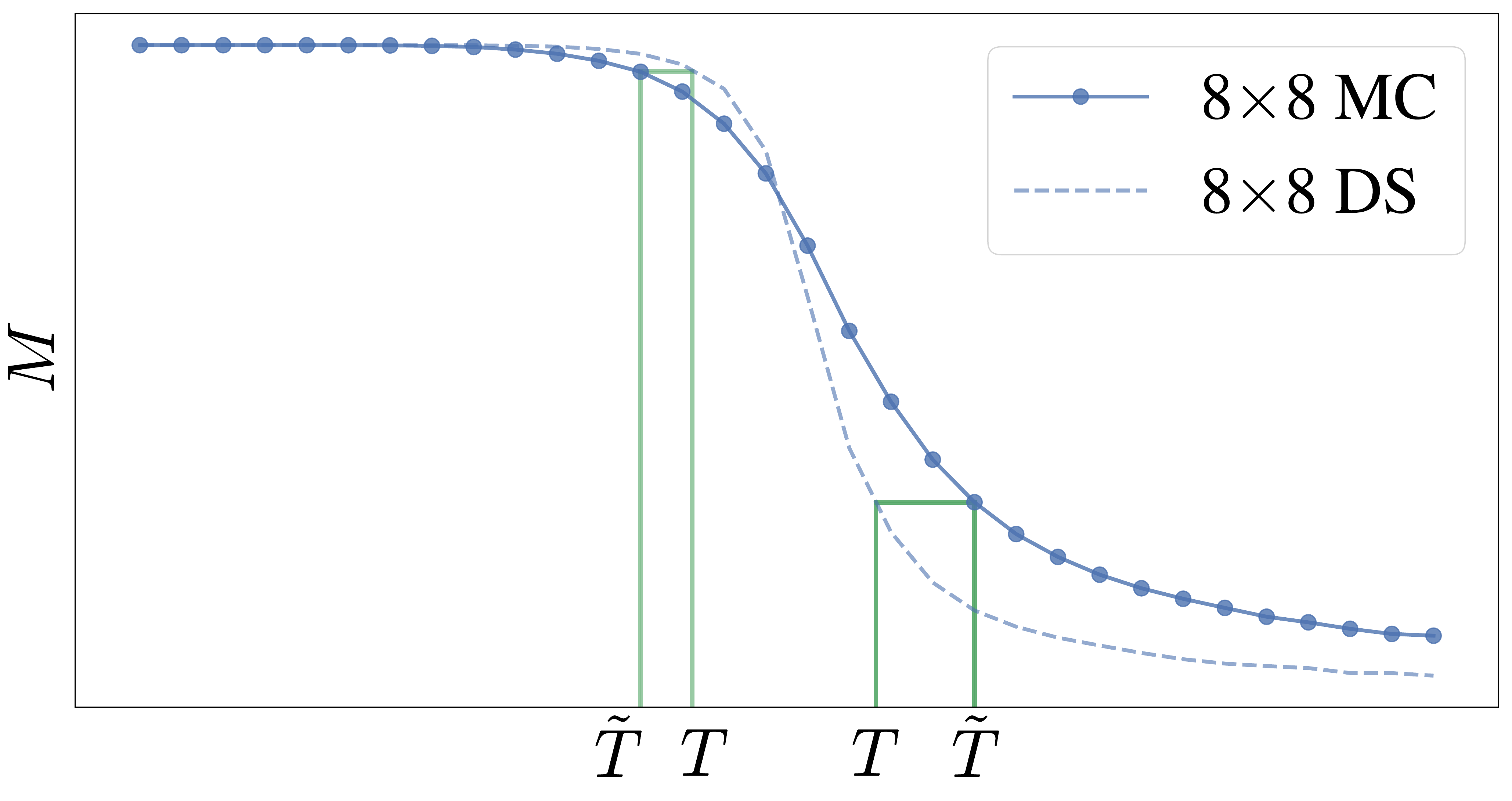}
    \caption{\label{fig:Ttransformation}The method we use to find the $T\rightarrow \tilde{T}$ rescaling. Note that SR data are not needed here.}
\end{figure}

As in the main text, let $\mathbf{s}_i\in \mathbb{Z}_2^{8\times 8}$ denote the configurations obtained upon applying $\mathcal{MR}$ to the $16\times 16$ Monte Carlo data $\boldsymbol{\sigma }_i$.
For the purpose of this Appendix we also sample $8\times 8$ Monte Carlo data denoted by $\boldsymbol{\tau }_i$.
To find the rescaling, we compare the magnetization calculated on $\boldsymbol{\tau }_i$ with that from $\mathbf{s}_i$.
We find the transformation $T\rightarrow \tilde{T}$ by requiring the corresponding $M(T)$ curves to collapse.
An easy way to do so is shown in Fig.~\ref{fig:Ttransformation}(a).
Starting from a point at temperature $\tilde{T}$ in the $\boldsymbol{\tau }_i$ curve, one moves horizontally towards the $\mathbf{s}_i$ curve and the intersection defines $T$.
By this construction, applying the rescaling to MC data, makes them collapse to downsampled (upon application of $\mathcal{MR}$) data of the same size.
Therefore this rescaling is equivalent to an approximation of $f^{-1}$ used to rescale SR data points in Fig.~\ref{fig:ups_real2D}.

We note that finding this rescaling is not related to the super-resolution procedure and it is not used in the critical exponent calculation, where we assume that we are at the fixed point of the $\mathcal{MR}$ transformation.

\section{Critical Exponents}\label{sec:app_expo}

\begin{figure}[htb]
    \centering
    \includegraphics[width=0.48\textwidth]{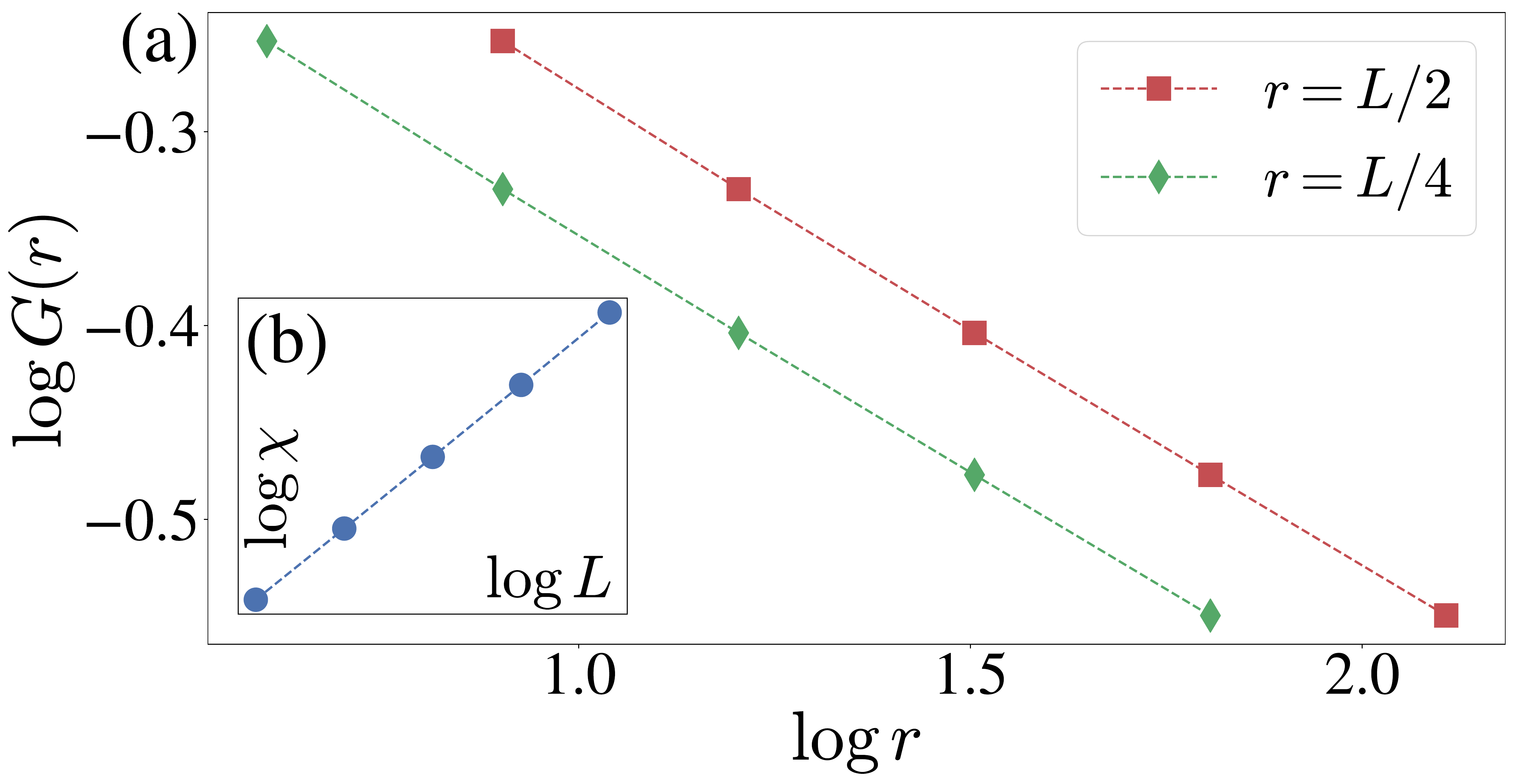}
    \caption{\label{fig:critical_linear}Scaling of the (a) two-point function and (b) susceptibility at criticality. The smallest size is calculated with Monte Carlo and the rest with repeated super-resolutions. Errors are typically around $10^{-3}$ and too small to show in this figure.}
\end{figure}

We calculate the values of susceptibility $\chi $ and the two point function $G(r)$ at every level of the repeated inverse RG procedure and give the corresponding log-log plots in Fig.~\ref{fig:critical_linear}.
The two-point function is calculated using two different values of the corresponding distance $r=L/4$ and $r=L/2$, leading to the two estimates $\eta _1$, $\eta _2$ for the anomalous dimension. To get an estimation of our method's error in the critical exponents, we repeat the training and critical exponent calculation 60 times for sizes up to $128\times 128$ and we give predicted values and percentage error in Table~\ref{tab:critical_exponents}.

The linearity of scaling can be confirmed from the correlation coefficient of the regression which differs from unity less than $10^{-4}$ in all cases and is also demonstrated in Fig.~\ref{fig:critical_linear}.
The linearity was confirmed for sizes up to $512\times 512$ using the repeated super-resolution procedure.
Moreover, the $y$-intercept in the $\chi $ fit agrees with the Monte Carlo calculation in Fig. 14 of Ref.~[\onlinecite{SandvikGuide}].
The most important result is that we achieve less than $2\%$ error from the theoretical value of the exponents in all cases.

\end{document}